\pdfoutput=1
\documentclass{JINST}

\usepackage{amsmath}
\usepackage{multirow}
\usepackage[amssymb,thinqspace,thinspace,Gray,textstyle]{SIunits}
\addunit{\ppm}{ppm}
\addunit{\Month}{month}
\addunit{\kV}{\kilo\volt}

\title{Next generation KATRIN high precision voltage divider for voltages up to 65kV}

\author{S. Bauer$^a$\thanks{Corresponding
author.}~, R. Berendes$^a$, F. Hochschulz$^a$\thanks{present address: 
Fraunhofer Institute for Microelectronic Circuits and Systems IMS
Finkenstr{\ss}e 61, 47057 Duisburg, Germany}, H.-W. Ortjohann$^a$, S. Rosendahl$^a$, T. Th\"ummler$^c$, M. Schmidt$^b$ and C. Weinheimer$^a$\\
\llap{$^a$}Institut f\"ur Kernphysik, Westf\"alische Wilhelms Universit\"at M\"unster,\\
  Wilhelm-Klemm-Stra{\ss}e 9, 48149 M\"unster, Germany\\
\llap{$^b$}Physikalisch Technische Bundesanstalt,\\
  Bundesallee 100, 38116 Braunschweig, Germany\\
\llap{$^c$}Institut f\"ur Kernphysik, Karlsruher Institut f\"ur Technologie,\\
  Hermann-von-Helmholtz-Platz 1, 76344 Eggenstein-Leopoldshafen, Germany\\
  E-mail: \email{s.bauer@uni-muenster.de}}

\abstract{The KATRIN (KArlsruhe TRItium Neutrino) experiment  aims to determine the mass of the electron antineutrino with a sensitivity of 200$\,$meV by precisely measuring the electron spectrum of the tritium beta decay. This will be done by the use of a retarding spectrometer of the MAC-E-Filter type. To achieve the desired sensitivity the stability of the retarding potential of -18.6$\,$kV has to be monitored with a precision of 3$\,$ppm over at least two months. Since this is not feasible with commercial devices, two ppm-class high voltage dividers were developed, following the concept of the standard divider for DC voltages of up to 100$\,$kV of the Physikalisch-Technische Bundesanstalt (PTB). In order to reach such high accuracies different effects have to be considered. The two most important ones are the temperature dependence of resistance and leakage currents, caused by insulators or corona discharges.
For the second divider improvements were made concerning the high-precision resistors and the thermal design of the divider. The improved resistors are the result of a cooperation with the manufacturer.  The design improvements, the investigation and the selection of the resistors, the built-in ripple probe and the calibrations at PTB will be reported here. The latter demonstrated a stability of about 0.1$\,$ppm/month over a period of two years.}

\begin{document}

\section{Introduction}

The KATRIN (KArlsruhe TRItium Neutrino) experiment \cite{KATRIN_design_report} aims to determine the mass of the electron antineutrino with a sensitivity of \unit{200}{\milli\electronvolt}. For that reason the electron spectrum of the tritium beta decay will be measured very precisely by the use of an electrostatic retarding spectrometer of the MAC-E-Filter (\textbf{M}agnetic \textbf{A}diabatic \textbf{C}ollimation with \textbf{E}lectrostatic filter) type \cite{MAC_E}. The stability of the retarding potential of \unit{-18.6}{\kilo\volt} has to be monitored with a precision of \unit{3}{\ppm}\footnote{\unit{1}{\ppm} = 1 part per million = 10$^{-6}$} over at least two months to achieve the desired sensitivity \cite{KATRIN_design_report}. 

Voltages in the kV regime or above are usually not directly measured but divided into lower voltages 
with the help of high voltage dividers. High-precision applications of high voltages in nuclear and particle physics are for example the
electric retarding spectrometers of MAC-E-Filter type of the former neutrino mass experiments at 
Mainz and Troitsk \cite{Mainz, Troitsk} or currently KATRIN \cite{KATRIN_design_report} 
or of experiments performing high-precision weak decay studies like aSpect or WITCH \cite{aSpect,witch}.
Other applications of precision high voltage in physics are electron coolers in storage rings \cite{ecooler}, the accelerating voltage of ions in collinear laser spectroscopy (e.g. \cite{Krieger}), or huge drift
 chambers or time projection chambers (e.g. \cite{alice_TPC}). Of course precision high voltage applications are very important also for technical applications. Recently a high-precision high voltage divider has been developed for high voltage direct current (HVDC) electric power transmission systems \cite{YiLi_Divider}.

\begin{figure}[h]
\centering
\includegraphics[width=0.7\textwidth]{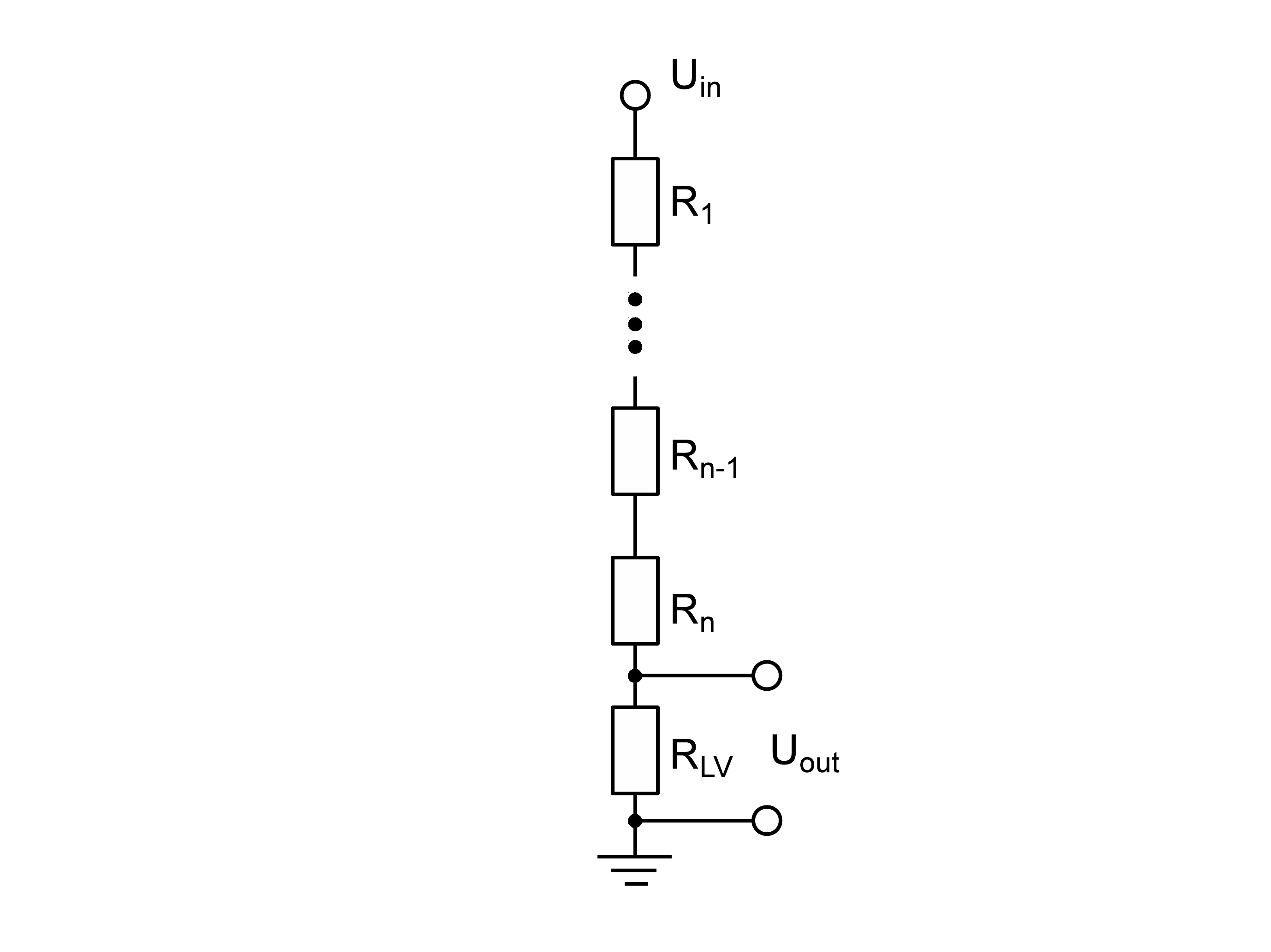}
\caption{Simple realisation of a high voltage divider. The high voltage resistance consists of a set of resistors $R_\mathrm{i},  \mathrm{i}=1 \dots n$ connected in series. The ratio between input voltage $U_\mathrm{in}$ and output voltage $U_\mathrm{out}$ is described by the scale factor given in equation (1.1).}
\label{fig::Divider_simple}
\end{figure}

The basic concept of a voltage divider is that two resistors are connected in series. For high voltage dividers the upper resistance $R_\mathrm{HV}$ usually consists of many resistors $R_\mathrm{i}$ connected in series in order to allow the requested maximum voltage (see figure \ref{fig::Divider_simple}).
Voltage dividers are usually characterized by a scale factor:

\begin{equation}
M := \frac{U_\mathrm{in}}{U_\mathrm{out}} = \frac{\sum\limits_{i=1}^n R_\mathrm{i} + R_\mathrm{LV}}{R_\mathrm{LV}}
\label{eq::scale_factor}
\end{equation}

The total resistance of a high voltage divider is a compromise between different properties which has to optimized for the application. In order to reduce the dissipated power of the divider and hence the influence of a temperature dependence of the resistors, the total resistance has to be as high as possible. On the other hand $R_{LV}$ will also increase for higher total resistance in order to meet the desired scale factor and hence the thermal voltage noise $U_\mathrm{n,eff} = \sqrt{4Rk_\mathrm{B}T\Delta f}$ over a frequency band $\Delta f$ will also increase. Secondly, the current which flows through the voltage divider must be significantly higher than the leakage currents. High-precision high voltage dividers typically use currents through the divider of about \unit{100}{\micro\ampere}. Therefore one has to find a suitable balance between high total resistance and a low $R_\mathrm{LV}$. 

For the use at the KATRIN experiment commercial high voltage dividers are not sufficient, since they do not surpass  a precision of the scale factor of 10$^{-4}$ to some 10$^{-5}$ with a significant thermal dependence and non-neglible long-term drifts. Only the Fluke 752A "Reference Divider" offers high accuracy of \unit{0.5}{\ppm} \cite{Flu752A} by the use of a self calibration procedure, but this device is limited to \unit{1}{\kV}. Therefore the only solution is a custom-made device. But also for self-development, it is very difficult to get into the area of the desired accuracy and stability, as there are different effects to be considered. One point is the thermal behaviour of the resistors, as they have a temperature coefficient of resistance (TCR). The TCR is expressed as absolute change of resistance with temperature  $\frac{\partial R}{\partial T}$ or as relative change $\frac{\partial R}{\partial T} R^{-1}$. An unbalanced change of all resistors directly leads to a change in the scale factor of the voltage divider. In order to counteract the thermal behaviour the resistors have to be selected according to their thermal characteristics and combined in a way that the overall thermal behaviour sums up to nearly zero temperature dependence of resistance. This can be further improved by an additional thermal stabilization of the voltage divider. The method we used for the selection is described in section \ref{Sec::ResSelection}.

As already mentioned, disturbances caused by the high voltage, such as corona discharges or leakage currents, can change the scale factor of a high voltage divider as well. In order to reduce leakage currents, insulators with high volume and surface resistance have to be used for the mounting structure of the resistors. A very suitable material for that purpose is PTFE\footnote{Polytetrafluoroethylene}. To prevent corona discharges, high field gradients must be avoided and the electric field inside the divider has to be homogeneous, with low field strengths. The first voltage divider which considered this was set up by Park using special shielded resistors \cite{Park}. By this shielding the potential difference surrounding the resistors is equal to the voltage drop over the resistor. Additionally a huge set of resistors was investigated with respect to their temperature coefficient of resistance. It was found that some showed a negative dependence and some a positive one. Therefore matched pairs were selected to achieve an average temperature coefficient close to zero. With the combination of both techniques, it was possible to set up a voltage divider ($R_\mathrm{total} = \unit{200}{\mega\ohm}$) with an accuracy of \unit{20}{\ppm} at \unit{100}{\kV} \cite{Park}. Knight and Martin reached an accuracy of \unit{5}{\ppm} ($1\sigma$) at \unit{100}{\kV} with a similar set-up \cite{Knight_Martin}. The major differences were a total resistance of \unit{1}{\giga\ohm} and additional corona shields which obtain their voltages from the  resistor chain itself. Two recent voltage dividers, with a set-up similar to that of Park, reach accuracies of \unit{30}{\ppm} (k=2)\footnote{$k=2$ means a coverage factor of 2 which is equivalent to a $\pm 2\, \sigma$ region} \cite{Thailand} and \unit{66}{\ppm} (k=2) \cite{Tuerkei}. Both have a total resistance of \unit{100}{\mega\ohm}. The most precise system based on the design of Park is the divider of the National Measurement Institute, Australia (NMIA) \cite{YiLi_Divider}. This divider reaches accuracies better than \unit{5}{\ppm} (k=2) at \unit{150}{\kV}. 

A slightly different approach was chosen for the MT100 which is the standard divider for DC voltages up to \unit{100}{\kV} of the PTB (Physikalisch Technische Bundesanstalt, national metrology institute of Germany) \cite{Marx}. This divider is equipped with 101 selected wire-wound resistors of \unit{10}{\mega\ohm} each. As it has already been done by Park, the resistors were also selected with respect to their temperature coefficient but with much higher accuracy. The resistors are arranged in a helix which is divided into five sections. These sections are separated by copper electrodes to shape the field inside the divider. Unlike the design of Park, the resistors itself are not individually enclosed by a shield, but as groups of 20 by the copper electrodes. The copper electrodes obtain their potential from an independent control divider in parallel to the high precision divider. The control divider also contains capacitances to protect the high precision divider against transient overload, when the voltage is switched on or off. Further the divider is enclosed by a steel vessel and filled with compressed SF$_6$\footnote{Sulfur hexafluoride}. This Faraday cage additionally screens the divider from external disturbances like RF noise and enables a temperature regulation of the system. A comparison between the MT100 and the NMIA divider showed deviations of less than \unit{2}{\ppm} \cite{Divider_comparison}.

Following the design of the MT100 two high precision voltage dividers were developed at the University of M\"unster in cooperation with the PTB. The first divider for voltages up to \unit{35}{\kV} (further referred to as K35) is described in \cite{K35}. The second one for voltages up to \unit{65}{\kV} (further referred to as K65) is a further development to higher voltages, an even higher precision and a better long-term stability w.r.t. the first divider (K35) and will be presented in this paper. For the KATRIN-experiment these dividers will precisely measure the retarding potential (\unit{-18.6}{\kilo\volt}) of the spectrometer over five years of measurement time in which the precision of \unit{3}{\ppm} is required for a period of two month \cite{KATRIN_design_report}. The K35 and K65 were also used to calibrate the acceleration voltage of the ISOLDE facility at CERN \cite{Krieger}.

\section{Design of the K65 divider}
\label{Design}

At the KATRIN-experiment the voltage under investigation is \unit{-18.6}{\kV}. Because the most common reference voltage sources have an output voltage of \unit{10}{\volt} and also the multimeter used (Agilent 3458A and Fluke 8508A) have the best performance in the \unit{10}{\volt} (Agilent 3458A) to \unit{20}{\volt} range (Fluke 8508A), the best scale factor would be around $M=2000:1$. For the K35 \cite{K35} it turned out that the best resistors, with respect to time stability and low TCR, were VISHAY Bulk Metal Foil\textregistered\ resistors (VHA518-11) \cite{Vishay}. The highest available resistance of these resistors is \unit{1.84}{\mega\ohm}. In order to meet the approved current of \unit{100}{\micro\ampere} at the nominal high voltage of \unit{-18.6}{\kV}, 100 resistors were chosen for the high voltage resistor chain leading to a total resistance of \unit{184}{\mega\ohm}. In combination with the low voltage resistance $R_\mathrm{LV}=\unit{93}{\kilo\ohm}$  a scale factor of $M$=1972:1 was obtained\footnote{The K35 has two further scale factors of 95:1 (after upgrade in 2009) and 3944:1}. The K35 already fulfils the KATRIN requirements but also left some space for improvements. 

The major improvements of the second high-precision KATRIN high voltage divider K65 are firstly the usage of even improved resistors exhibiting a very low TCR by  matching positive and negative temperature coefficients inside the resistor itself and an improved long-term stability (section \ref{Sec::ResSelection}). 
These newly developed resistors are limited in resistance to \unit{880}{\kilo\ohm}.  To reach a total resistance similar to that of the K35, a much higher number of resistors is requested leading as the second advantage to a lower thermal load per resistor at the nominal high voltage of \unit{-18.6}{\kV}. The larger number of resistors and a larger stainless steel vessel provided an additional feature:  The maximum operating voltage of the high voltage divider K65  is increased to \unit{65}{\kV} expanding the field of possible applications, e.g. for further calibrations at the ISOLDE facility \cite{Krieger}.
Thirdly the thermal distribution inside the divider is improved (section \ref{Sec::ThermDistr}).

 We decided to use 165 resistors of \unit{880}{\kilo\ohm}. 
 In principle a higher number of resistors would have been possible but would have significantly complicated and increased the size of the mechanical set-up. 
 The 165 resistors were selected from a total of 194 resistors to obtain the smallest possible TCR.
The  load per resistor at the nominal high voltage of \unit{-18.6}{\kV} was
 reduced from \unit{19}{\milli\watt} (K35) to \unit{14}{\milli\watt}. 
The chosen set of resistors leads to a total resistance of  \unit{$\approx 147$}{\mega\ohm} including the resistors of the low voltage section summarized as $R_\mathrm{LV}$. The resistors are arranged in pairs or in small groups in such a way that one resistor with positive TCR is placed next to one with a negative TCR of the same absolute value. By this it is ensured that both resistors have the same ambient temperature and their thermal behaviour adds to nearly zero, as presented in section \ref{Sec::ResSelection}.

\begin{figure}[!!t]
    \includegraphics[width=\textwidth]{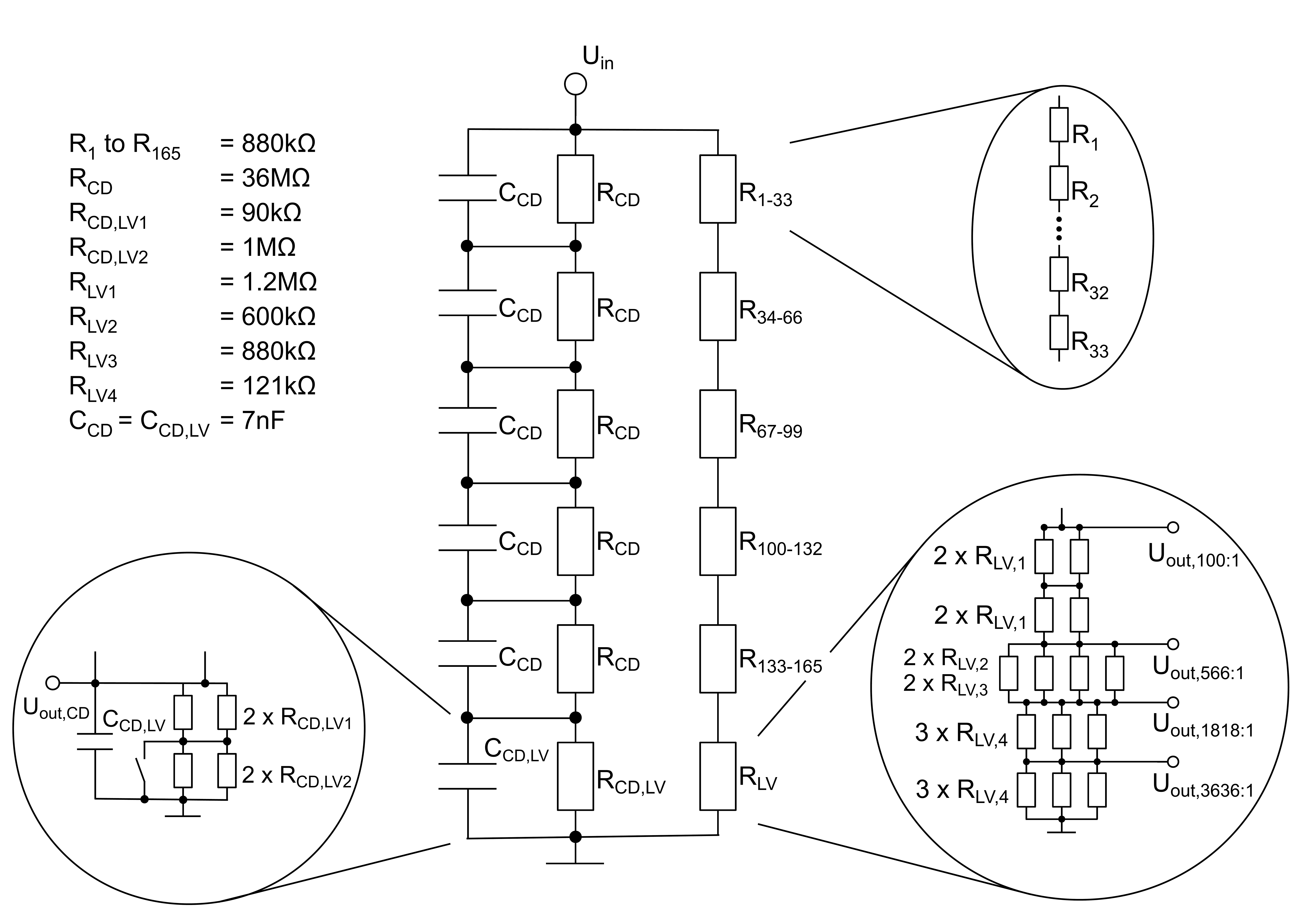} 
\caption{The equivalent circuit of the K65 divider. The low voltage part of the primary divider is shown in detail in the lower right inset. The tap of the built-in ripple probe including the additional resistor $R_\mathrm{CD,LV2}$ is shown in the lower left inset. The corresponding values are listed in the upper left part of the picture.}
\label{fig:Teiler_einfach}
\end{figure}

The high voltage resistor chain is followed by resistors of the same type but lower values in the low voltage section providing different low voltage outputs (fig. \ref{fig:Teiler_einfach}). The nominal values of the corresponding scale factors are 3636:1, 1818:1, 566:1 and 100:1 and the exact values are determined in measurements at the PTB (see sec. \ref{Sec::PTB}) and at the Institut f\"ur Kernphysik, University of M\"unster, Germany. The scale factor to be used for standard measurements at KATRIN is the 1818:1 scale factor. 
For high voltage measurements above $| U | = $\unit{36}{\kV} the 3636:1 scale factor is used, because the high precision measurement range of the multimeter (Fluke 8508A) is limited to \unit{20}{\volt}. Additionally high resolution measurements for voltages up to \unit{11}{\kV} are possible because of the 566:1 scale factor. The 100:1 scale factor is necessary for calibration purposes, since the commonly used reference divider Fluke 752A can measure voltages up to \unit{1}{\kV} with an accuracy of \unit{0.5}{\ppm} \cite{Flu752A}. 

\begin{figure}[!!tbp]
\begin{center}
\includegraphics[width=0.44\linewidth]{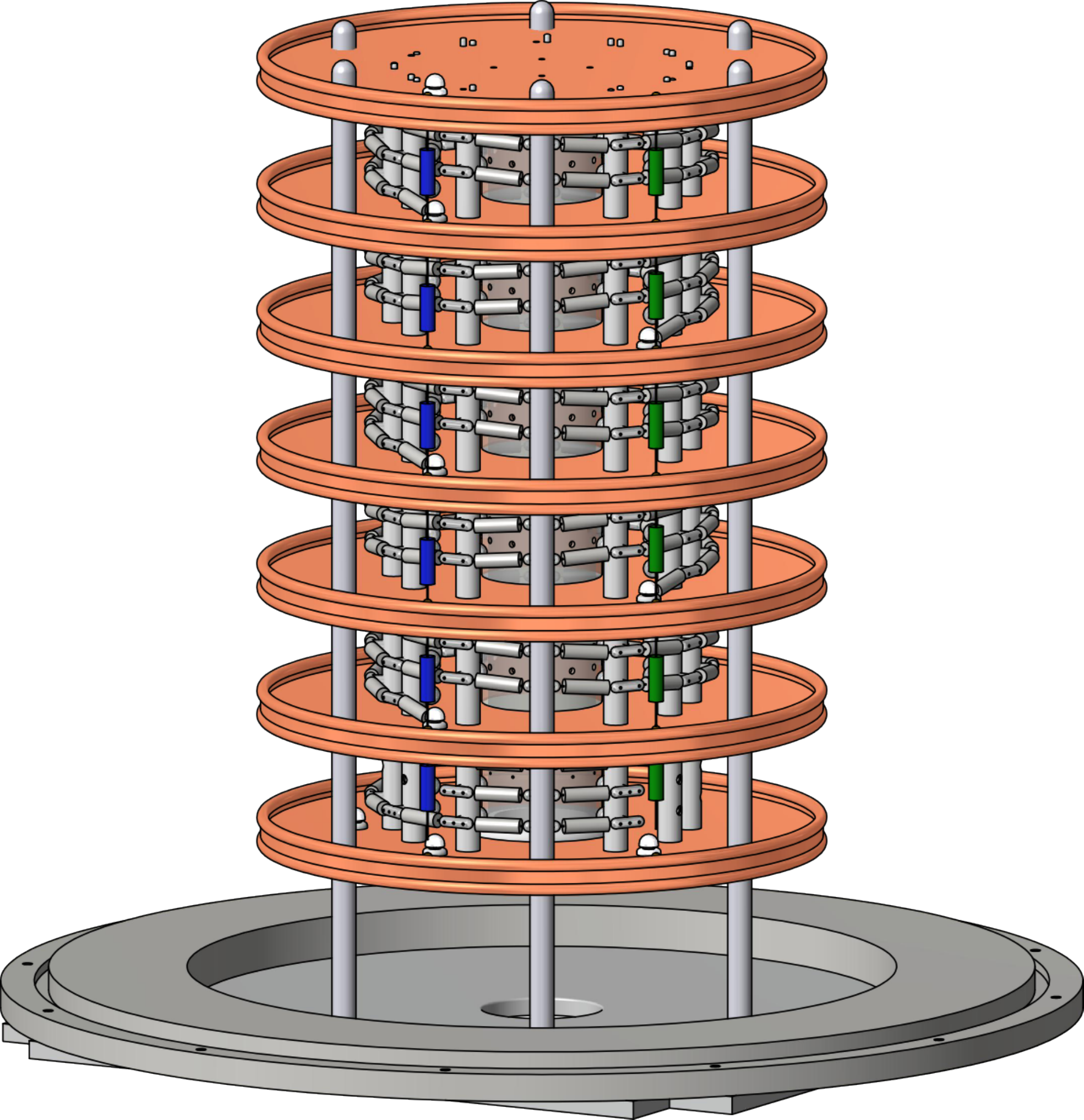} 
\includegraphics[width=0.55\linewidth]{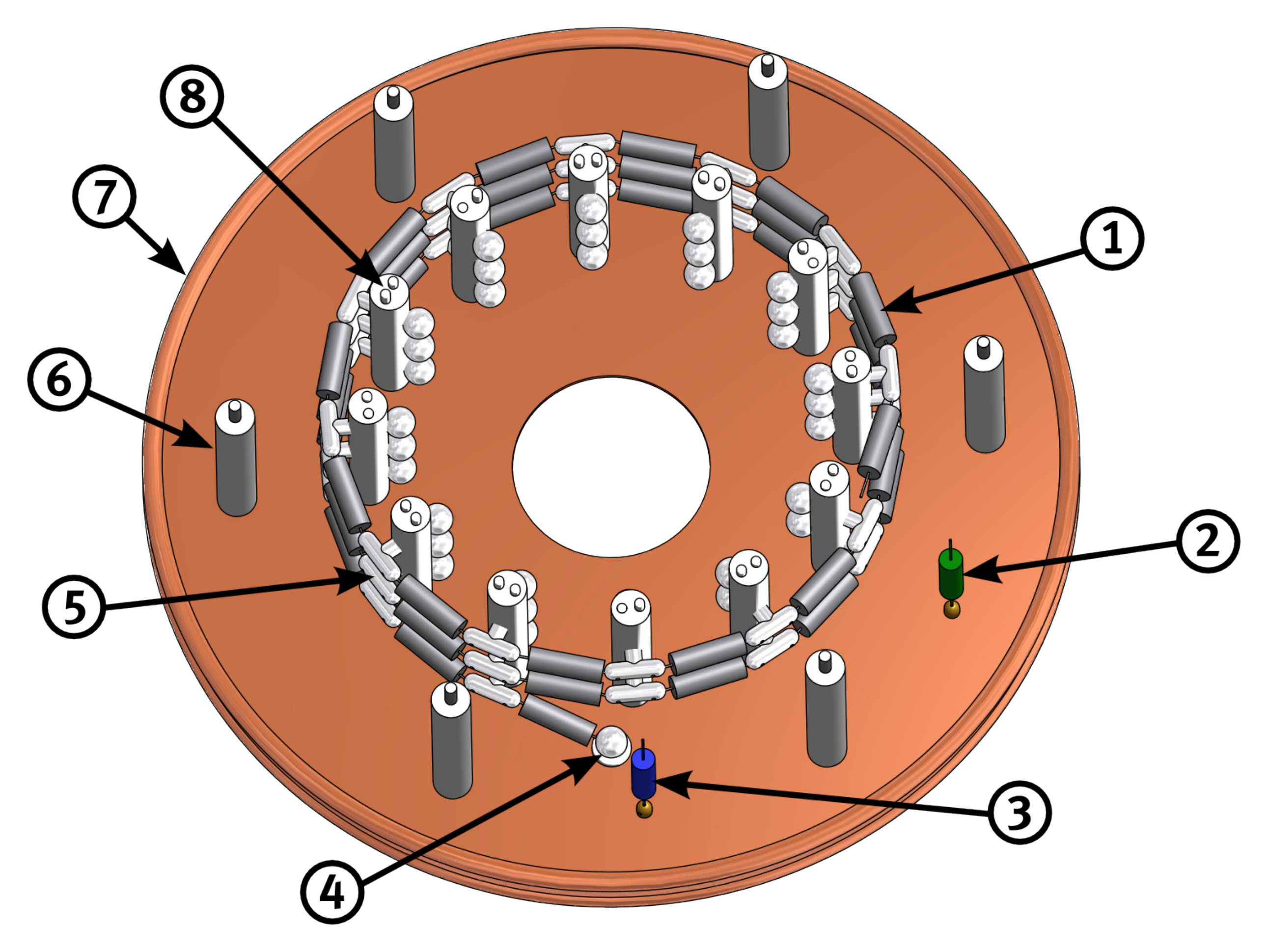}
\caption{CAD model of the high voltage divider and detailed view of one high voltage resistor section. The left picture shows the complete divider without the stainless steel vessel. The upper five sections house the 165 precision resistors and the lowermost section contains the low voltage resistors, which provide the different scale factors. In the center the PMMA tube of the heating/cooling system is shown in gray. The right picture shows one section of the high voltage resistor in detail. Each section contains the following parts: \textcircled{1} 33 precision resistors, \textcircled{2} one high voltage capacitor and \textcircled{3} one high voltage resistor of the control divider, \textcircled{4} insulated feedthrough for the precision resistor chain, \textcircled{5} Nickel-plated brass mountings, \textcircled{6} POM supports, \textcircled{7} Copper electrode, \textcircled{8} PTFE supports. }
\label{fig:Aufbau}
\end{center}
\end{figure}

At the K65 the high voltage resistor chain is divided into five sections of 33 resistors each, separated by copper electrodes which form a driven guard (see fig. \ref{fig:Aufbau}). The high precision resistors are arranged in a helix which changes direction  after each section (see fig. \ref{fig:Aufbau}) to minimize the inductance of the resistor chain. All resistors are mounted on nickel-plated brass mountings. These mountings are fixed on PTFE supports, which offer high-insulating resistances. The copper electrodes are mounted on supports made from POM\footnote{Polyoxymethylene} (see fig \ref{fig:Aufbau}).

A second resistive-capacitive divider chain is connected parallel to the high precision divider chain. This divider is on the one hand the control divider for the driven guard and on the other hand the capacitive chain to protect the high precision divider against transient overloads. It also enables low precision monitoring of the applied high voltage and exhibits a built-in ripple probe (see section \ref{Sec::Ripple_Probe}). A circuit diagram of the K65 divider is shown in figure \ref{fig:Teiler_einfach}.

The complete set-up is stored in a stainless steel vessel which works as a Faraday cage to reduce the influence of external disturbances like RF signals. Furthermore the vessel  provides a sealed surrounding for an atmosphere of dry nitrogen gas at a slight overpressure of \unit{70}{\milli\bbar} allowing a temperature stabilisation and avoiding humidity-driven leakage currents . 

%

The temperature stabilisation is realized by a water circulation heating-cooling system. Thus the voltage divider is decoupled from the electrical components of the heating-cooling system to avoid possible noise from external devices. The main components outside the high voltage divider vessel are a thermoelectric cooler/heater with a heat exchanger and a water pump. Inside the vessel a second heat exchanger connects the temperature of the water with the one of the dry nitrogen atmosphere. A fan transports the nitrogen through a tube made from PMMA\footnote{Polymethylmethacrylate, also known as acrylic glass under the trade name Plexiglass} via bores in the tube wall to the resistors. The bores were optimized to control the thermal distribution inside the divider (see section \ref{Sec::ThermDistr}). The control of the system is done by a PID\footnote{Proportional-integral-derivative controller}  controller, realized with the software LabView. The humidity inside the stainless steel vessel is monitored by a humidity sensor. If the humidity is higher than 30\% the nitrogen gas is exchanged. This sensor can also be used for detecting leaks in the water circuit, since the relative humidity will significantly increase if a leakage occurs within the vessel.

\section{Selection of resistors}
\label{Sec::ResSelection}

\subsection*{Resistor technology}

For the K65 high voltage divider the same type of resistors are used as for the K35, but in an improved version. The improvements were realized through a close cooperation between the University of M\"unster and the company VISHAY. VISHAY's ``Hermetically Sealed High Precision Bulk Metal Foil\textregistered\ Technology Resistors'' consists of several chips mounted inside a tinned brass cylinder, which is filled with oil for a better heat transport from the chips to the cylinder. Each chip is made out of a ceramic substrate with a meander-type metal film on it. Since both materials have different thermal expansion coefficients, the metal film reacts on the mechanical stress due to temperature changes by a mechanical deformation which is a wanted feature. By designing the chip the deformation under temperature change can be controlled and thus temperature coefficients of resistances (TCR) of both signs can be realized by the manufacturing process. University of M\"unster's task within that cooperation was to characterize with its dedicated high-precision set-up resistors which were produced at VISHAY with certain production parameters. The aim was to build resistors consisting of chips with both positive and negative TCRs compensating each other such, that the overall TCR becomes close to zero but with a defined sign. 
It should be mentioned that this compensation holds for a given operation temperature, which was in our case \unit{25}{\celsius}.

The new manufacturing process in combination with the required ultra-low TCR was only possible for resistors of up to \unit{880}{\kilo\ohm}. This fits in well with the already proposed increased number of resistors. The set of resistors of the new type VHA518-11 were ordered to have the same amount of slightly positive and negative TCRs.
These resistors were additionally treated by VISHAY with a special procedure ("pre-ageing") to improve the long term stability.

\subsection*{Selection procedure}

\begin{figure}[!!htbp]
\centering
\includegraphics[width=0.5\textwidth]{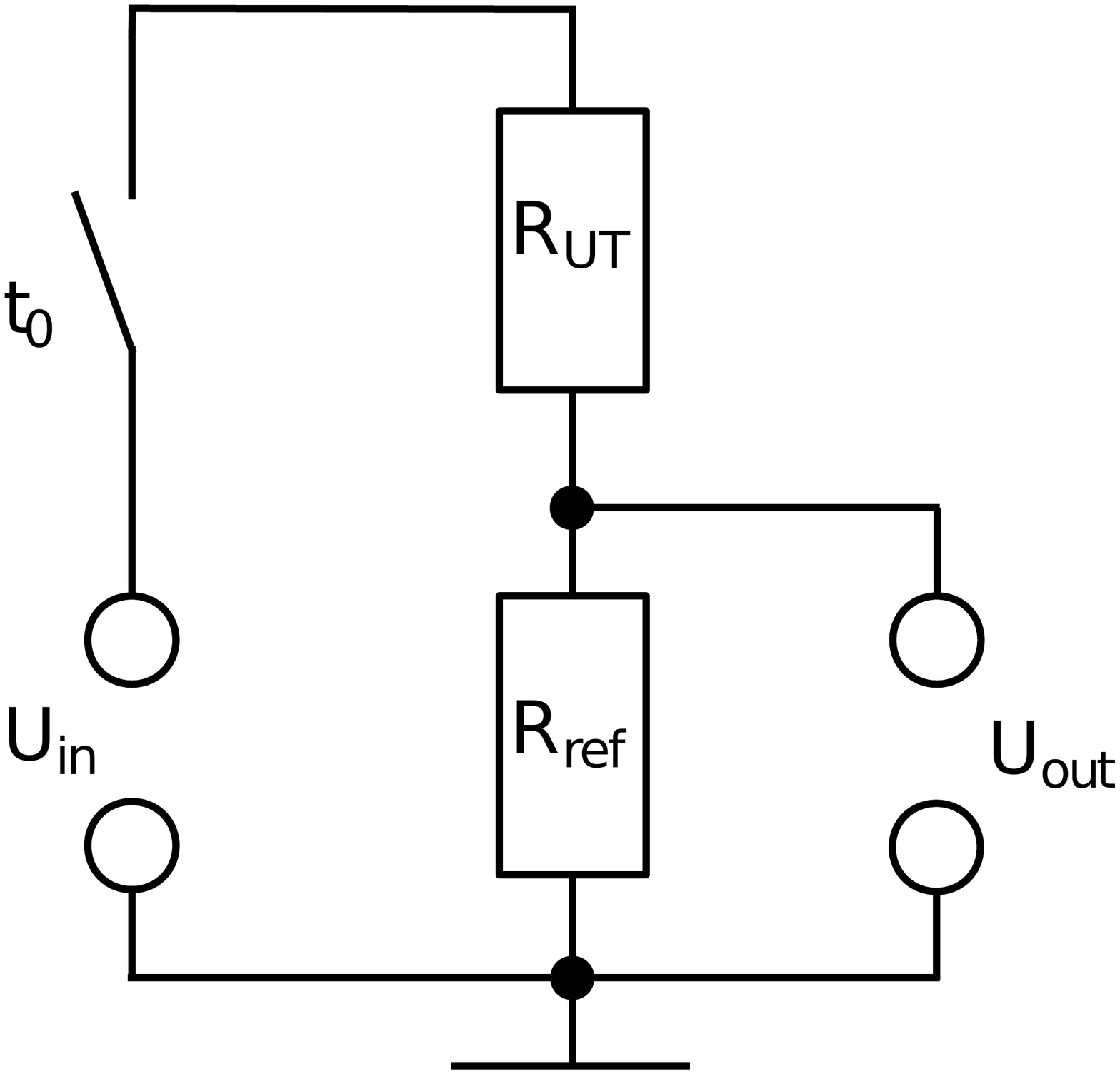}
\caption{Circuit of the voltage divider arrangement to measure the WUD. During measurement this divider is placed in a temperature stabilized box at \unit{(25.0$\pm$0.1)}{\celsius}.}
\label{fig::Test_divider_WUD}
\end{figure}

To determine the relative change of resistance with temperature one can chose different methods. 
One can measure the temperature coefficient of resistance (TCR) directly by changing the temperature of the resistor while a constant voltage is applied. Another method is to determine the warm-up drift (WUD) by measuring the change of resistance over time after a defined voltage is applied, while the ambient temperature is kept constant. For both methods the resistor under test $R_\mathrm{UT}$ is connected in series with a reference resistor $R_\mathrm{ref}$ to form a simple divider circuit (see figure \ref{fig::Test_divider_WUD}). We used a reference resistor  with $R_\mathrm{ref} = $\unit{36.8}{\kilo\ohm}, thus the thermal load of this resistor is negligible. In order to not compromise the measurement precision it was also a high-precision Bulk Metal Foil\textregistered\ resistor from VISHAY.

The investigation of a representative set of resistors proved a strong correlation between WUD and TCR.
Because the measurements of the WUD are much more simple to carry out, this method was chosen for the resistor selection of both KATRIN dividers \cite{K35,Hochschulz2008}. A further advantage is, that the resistors are tested under conditions of operation. During operation the resistors will have a stable ambient temperature and a temperature gradient from the inside to the outside. This gradient is caused by the  heating of the resistor due to the power dissipation in combination with the limited thermal conductance of the oil inside the resistor. 

The input voltage was delivered by a Fluke 5720A Calibrator and the voltage drop ($U_\mathrm{out}$) over the reference resistor was measured with an Agilent 3458A voltmeter. Both devices were granted a warm-up time of at least four hours to achieve the maximum stability. Stable conditions during the WUD measurements are realized by placing the resistors in a temperature stabilized box at the default
operation temperature of the resistors of \unit{(25.0$\pm$0.1)}{\celsius}. This box also acts as Faraday cage and screens the set-up against external influences. 

To measure the WUD the following procedure was chosen:
After applying \unit{255}{\volt} at the time t$_0$ the voltage drop over the reference resistor was measured every \unit{10}{\second} over a period of \unit{30}{\minute}. The relative change of the measured voltage 
is the relative change of resistance 
 times a small correction factor considering the reference resistor:
\begin{equation}
\frac{\Delta U_\mathrm{out}}{U_\mathrm{out}} = \frac{\Delta R_\mathrm{UT}}{R_\mathrm{UT}} 
\cdot \frac{R_\mathrm{UT}}{R_\mathrm{UT}+R_\mathrm{ref}} 
:= \alpha_\mathrm{WUD} \cdot \frac{R_\mathrm{UT}}{R_\mathrm{UT}+R_\mathrm{ref}}
\end{equation} 

\begin{figure}[!!!htbp]
\begin{center}
\includegraphics[width=0.85\linewidth]{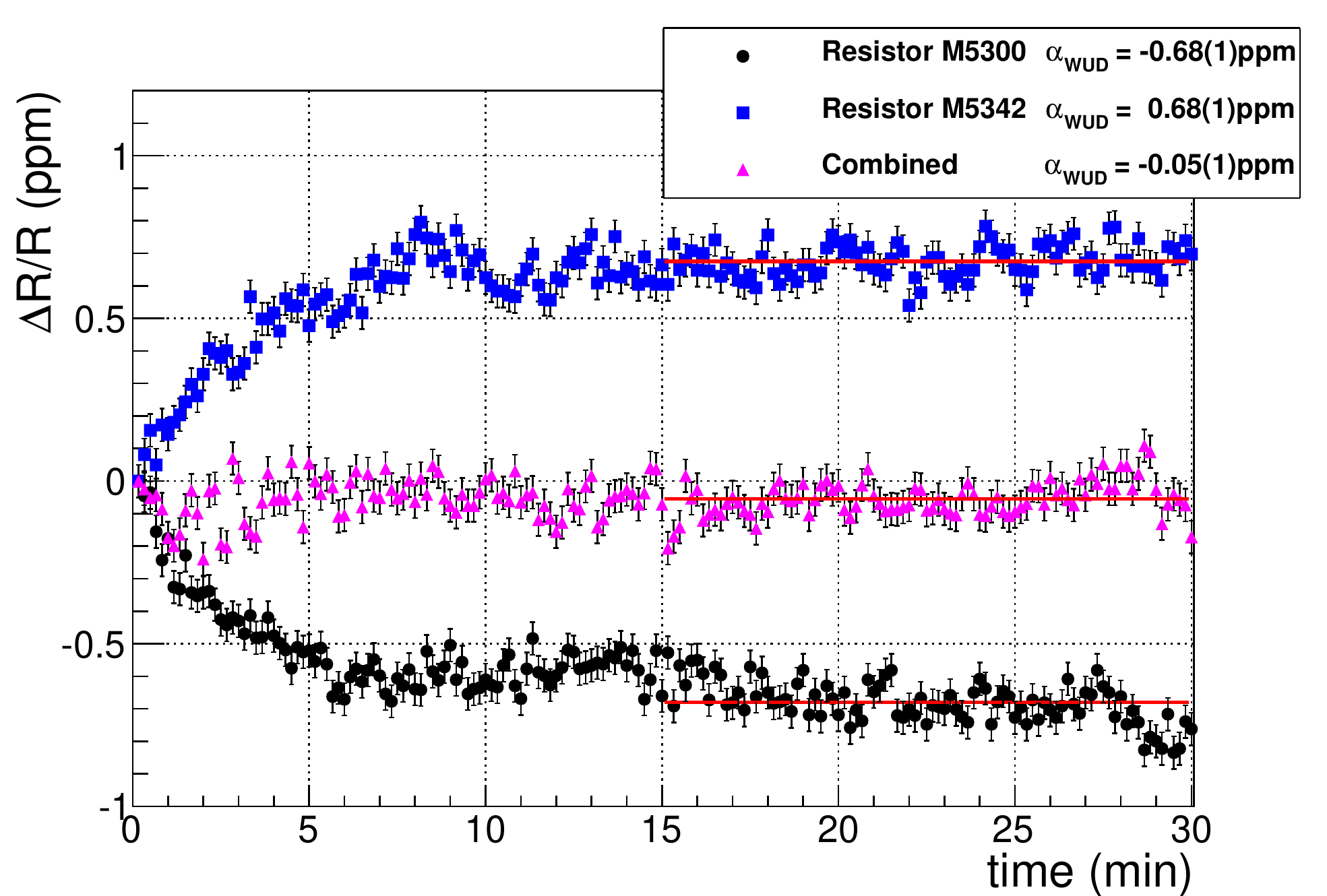} 
\caption{The warm-up drift $\alpha_\mathrm{WUD}$ for $U_\mathrm{in}= \unit{255.2}{\volt}$ for the resistors named M5300 and M5342 is shown. The triangles shows the residual warm-up drift $\alpha_\mathrm{WUD}$ for  $U_\mathrm{in}= \unit{510.4}{\volt}$ after  both resistors were connected in series. The red lines denotes the constant fit to the last \unit{15}{\minute}. The legend exhibits the fits and its statistical uncertainties.}
\label{fig:WUD}
\end{center}
\end{figure}

\begin{figure}[!!!htbp]
\begin{center}
\includegraphics[width=0.85\linewidth]{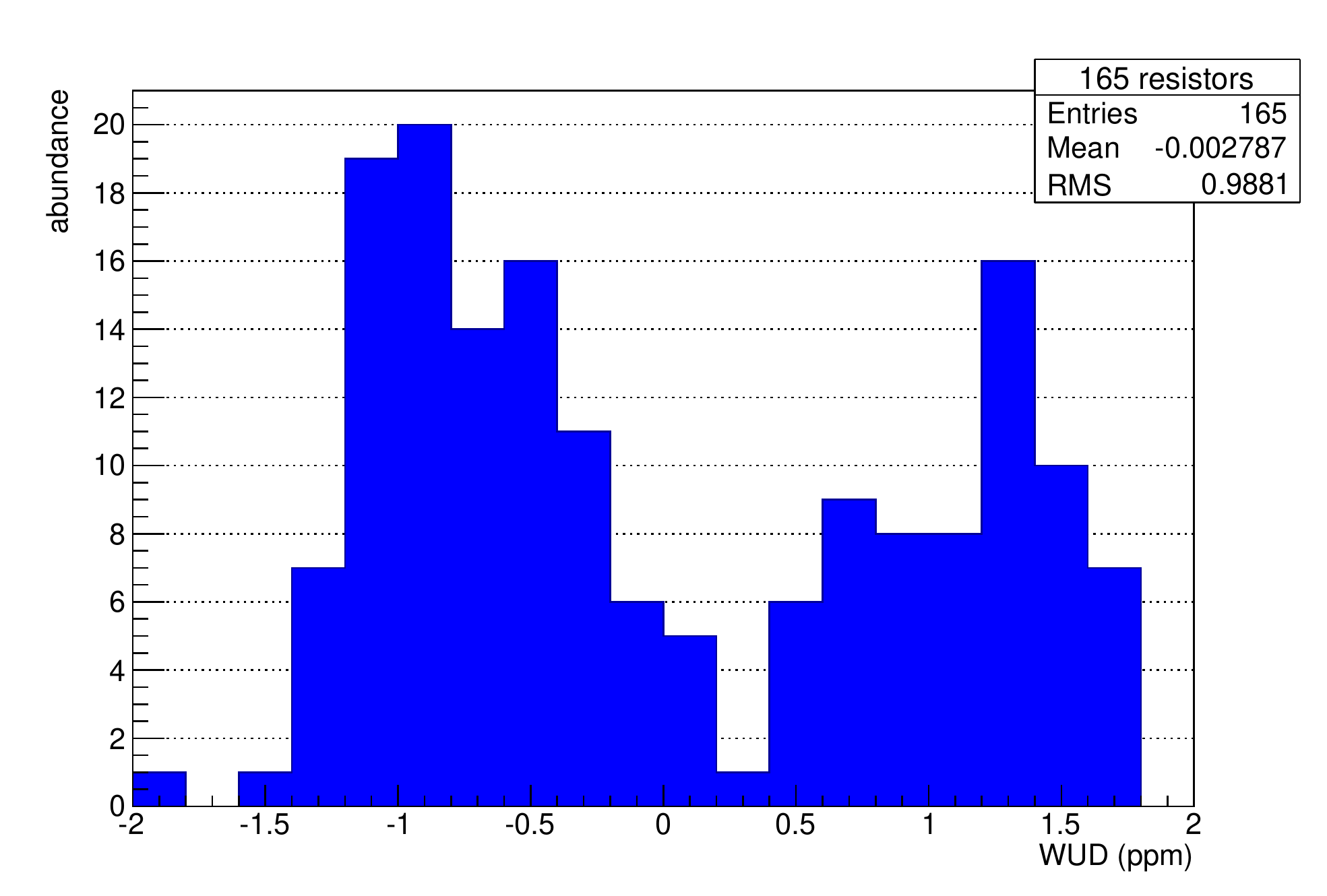} 
\caption{Results of the 165 selected resistor for the K65 divider. They were produced to have the same amount of resistors with a positive and negative $\alpha_\mathrm{WUD}$. The $\alpha_\mathrm{WUD}$ of all resistors sums up to \unit{-0.46}{\ppm} which corresponds to a mean $\alpha_\mathrm{WUD}$ of \unit{-0.003}{\ppm} per resistor.}
\label{fig:ResSelec}
\end{center}
\end{figure}

 This change over time is the WUD coefficient $\alpha_\mathrm{WUD}$ (see fig. \ref{fig:WUD}). The WUD coefficient was determined by fitting a constant to the data of the last \unit{15}{\minute}. 
 The short-term fluctuations of the measurement equipment is not well known, since the manufacturer usually provides the long-term drifts for 8 or 24 hours. Therefore we determine the short-term fluctuations
 of our set-up by requesting that the reduced $\chi^2$ of the fits are about one yielding a measurement uncertainty of the relative change of resistance of \unit{0.05}{\ppm} (see error bars in  fig. \ref{fig:WUD}). 
 The results of the resistor selection is shown in figure \ref{fig:ResSelec}. This figure also demonstrates that VISHAY indeed delivered resistors with positive and negative $\alpha_\mathrm{WUD}$ as requested.

\clearpage

\section{Optimization of thermal distribution} 
\label{Sec::ThermDistr}

In order to find possible improvements for the thermal design, the temperature distribution inside the K65 divider was simulated with EFDLab \cite{Hochschulz2008}. Following the optimal parameters found by this simulations the design for the temperature regulation and the distribution tube inside the divider was made. The major differences in the thermal design compared to the K35 are four additional bores in the topmost electrode section to improve the nitrogen flow in the top part of the vessel and the low voltage resistors also get a reduced flow of thermalized nitrogen gas.

To verify the simulations, a test set-up with the final mechanical structure but with common \unit{2.2}{\kilo\ohm} \unit{1/4}{\watt} resistors was made. By this the same thermal load as at high voltages can be simulated but at much lower voltages. Hence it was possible to simulate the thermal load of \unit{40}{\kilo\volt} by applying \unit{2}{\kilo\volt}. This low voltage makes it possible to place temperature sensors (type: PT100) in every section of the divider. All together eight sensors were installed at different positions inside the divider (two sensors in the low voltage section, one sensor in the top, middle and bottom plane each, one above the top electrode and one below the heat exchanger).

\begin{figure}[!!bp]
\centering
       \includegraphics[width=0.9\textwidth]{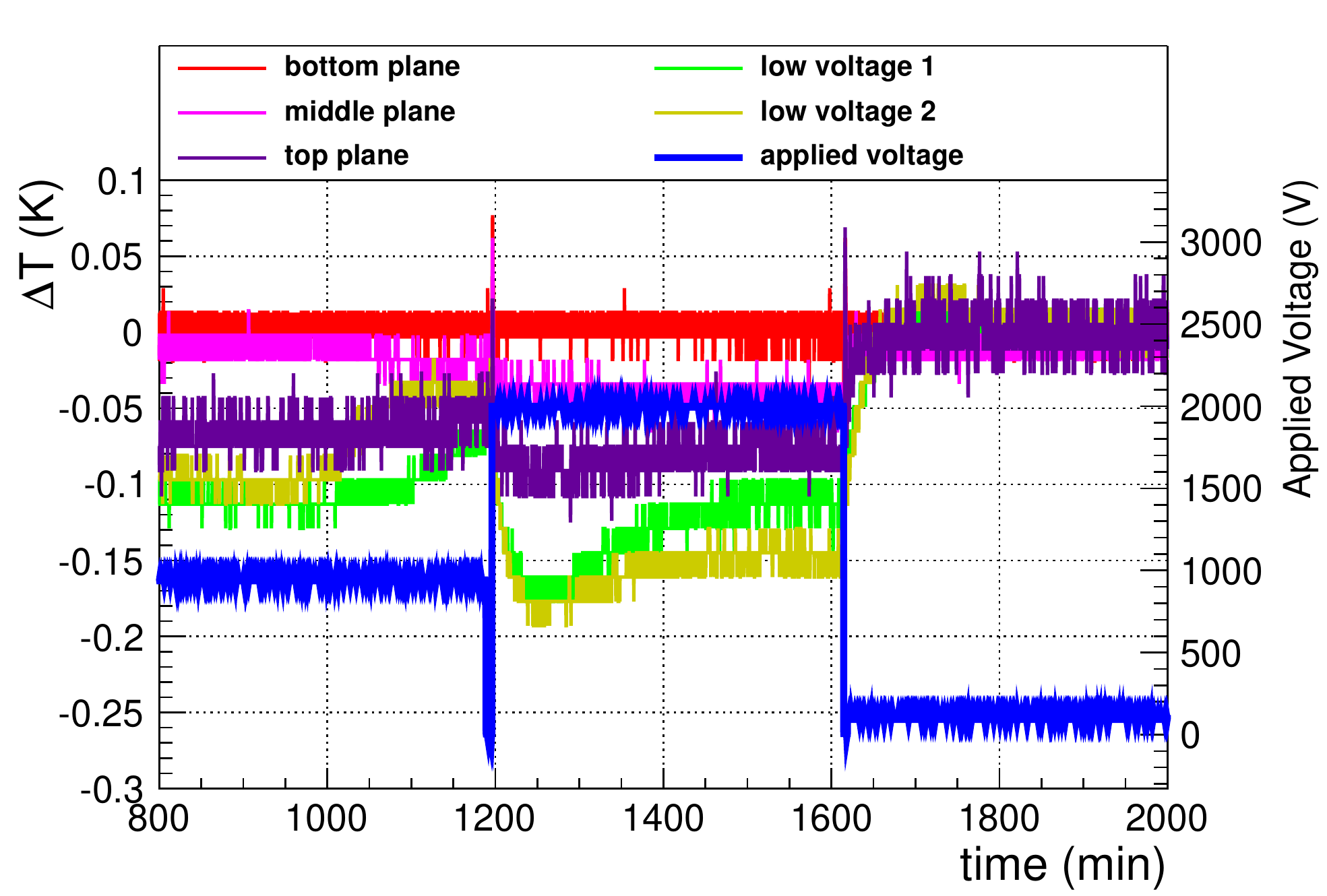}
\caption{
Relative change in temperature inside the divider with the final PMMA tube. The plot shows the reading of five temperature sensors inside the divider while the applied voltage (blue curve) was switched
from 1 to \unit{2}{\kilo\volt} and back to \unit{0}{\kV}. During the measurements the control variable for the PID was the temperature sensor at the bottom plane. All sensor values were normalized to the their mean values of the last \unit{10}{\minute}. The change in temperature is \unit{<0.2}{\kelvin} which was achieved by a new design of the PMMA tube.}
\label{fig:Temp_dist}
\end{figure}

After applying \unit{2}{\kilo\volt} to the divider the temperature in the low voltage area decreases by \unit{0.6}{\kelvin}. The reason for this was a not optimal distribution of holes in the PMMA tube. The high cooling power, caused by the warming of the resistors in the high voltage part, combined with the low dissipated heat in the low voltage area leads to this drop in temperature there. The thermal distribution was improved by a new tube. This one has 36 holes with a diameter of \unit{5}{\milli\meter} in each section of the high voltage resistor and 12 holes with a diameter of \unit{2}{\milli\meter} in the low voltage section. With this new tube the temperature drop is reduced to less then \unit{0.2}{\kelvin} (figure \ref{fig:Temp_dist}). This is acceptable because of a good TCR matching of the resistors in the low voltage area.

\section{Built-in ripple probe}
\label{Sec::Ripple_Probe}

One additional feature of the K65 divider is the built in ripple probe. It can be used to investigate noise on DC high voltages. This can be done during the ppm precise measurement of high DC voltages of up to \unit{65}{\kilo\volt}. The ideal case of a ripple probe is a high-pass filter composed of a capacitor (C) and a resistor (R) connected in series. For frequencies much higher than the cut-off frequency ($f_\mathrm{c}=\frac{1}{2\pi RC}$) the scale factor of such a probe is equal to 1. The K65 uses the control divider tap as ripple probe with the consequence that for low frequencies the scale factor will be dominated by the resistive part of the control divider ("low precision HV-divider") and for high frequencies by the scale factor of the capacitive part ("capacitive divider"). This capacitive divider causes the scale factor for high frequencies to be around 6 instead of 1. For measurements with the ripple probe two additional resistors ($R_\mathrm{CD,LV2}=\unit{500}{\kilo\ohm}$, see figure \ref{fig:Teiler_einfach}) can be switched into the low voltage resistor chain of the control divider chain to reduce the cut-off-frequency to approx. \unit{60}{\hertz}. This allows to investigate noise also in the regime of the line frequency (\unit{50-60}{\hertz}). 

\begin{figure}[!!!!!htbp]
\begin{center}
\includegraphics[width=0.95\linewidth]{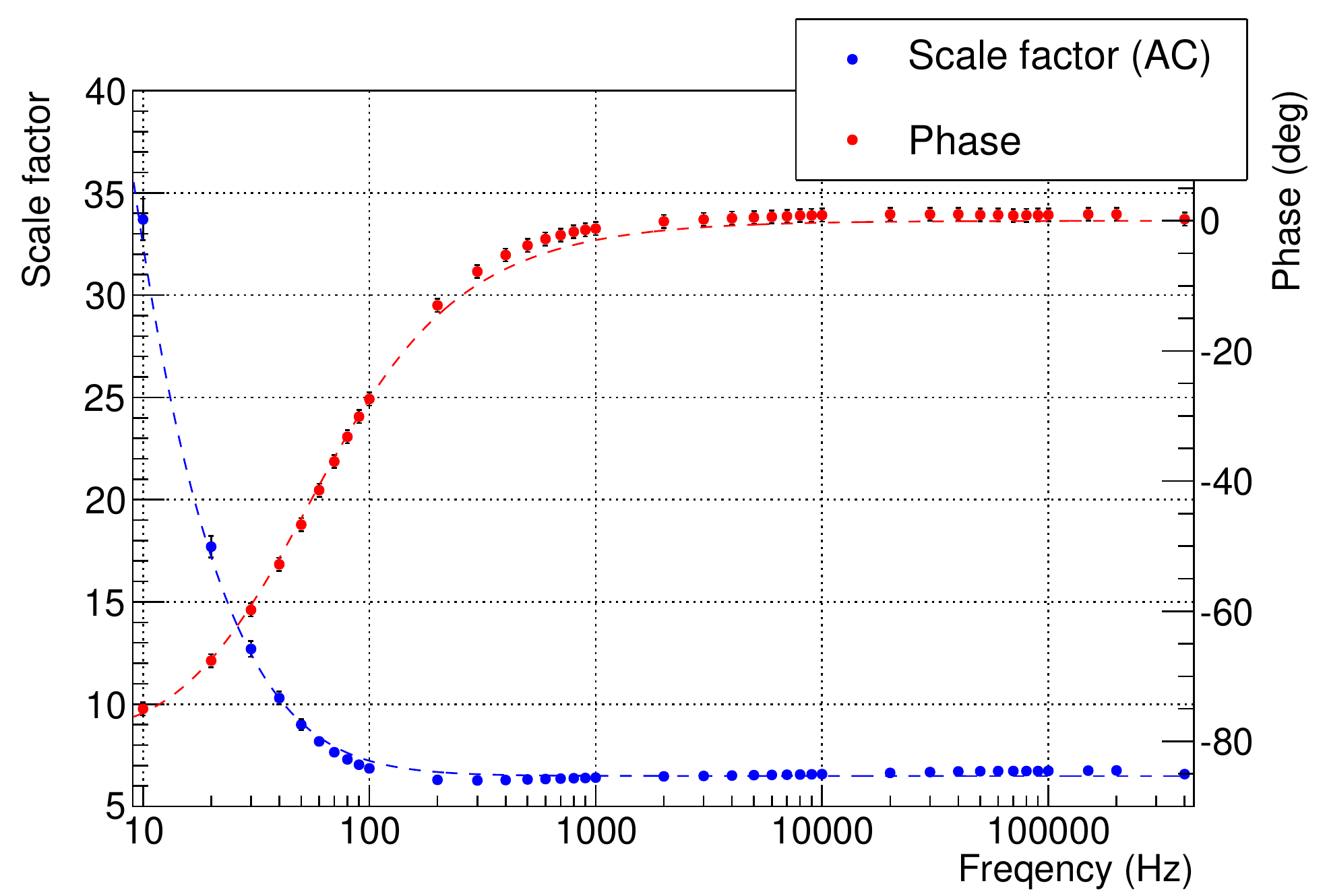}
\caption{Measurement of the scale factor (blue points) and phase (red points) of the frequency response of the built-in ripple probe with the two additional resistors $R_\mathrm{CD,LV2}$ switched into the low voltage resistor chain.  A fit to the data points is denoted as dashed lines.}
\label{fig:Teiler_ripple}
\end{center}
\end{figure}

Figure \ref{fig:Teiler_ripple} shows a measurement of the 	frequency response characteristic and phase difference of the built-in ripple probe. For this measurement the AC voltage of \unit{3}{V}$_\mathrm{RMS}$ was delivered by a Fluke 5720A Calibrator and the output voltage was measured with a National Instruments PXI 5922 digitizer. The frequency was varied from \unit{10}{\hertz} to \unit{400}{\kilo\hertz}. 

In order to find a simple description of the ripple probe the high voltage and the low voltage part of the control divider is combined to one equivalent resistance and one equivalent capacitance each:

\begin{align}
R^\mathrm{HV}_\mathrm{CD} &= 5\,R_\mathrm{CD} & C^\mathrm{HV}_\mathrm{CD} &= \frac{C_\mathrm{CD}}{5}\\
R'_\mathrm{CD,LV} &= \frac{R_\mathrm{CD,LV1}}{2} + \frac{R_\mathrm{CD,LV2}}{2} & C_\mathrm{CD,LV} &= C_\mathrm{CD}
\end{align}

In figure \ref{fig:Teiler_ripple} the theoretical description of this system was fitted to the measurements. For $R'_\mathrm{CD,LV}$ and $C_\mathrm{CD,LV}$ the influence caused by the impedance of the measurement device is considered in the theoretical description. 
Since the fit is only sensitive to the product of $R \cdot C$ the values for $R^\mathrm{HV}_\mathrm{CD}$ and $R'_\mathrm{CD,LV}$ were fixed to the values measured with a Fluke 8508A. 
The capacitances $C'_\mathrm{CD}=\unit{1.4(2)}{\nano\farad}$ and 
$C_\mathrm{CD,LV}=\unit{7.1(2)}{\nano\farad}$ were obtained by the fit. 
This is in nice agreement with the measured values of $C^\mathrm{HV}_\mathrm{CD,meas}=\unit{1.4(1)}{\nano\farad}$ and $C_\mathrm{CD,LV,meas}=\unit{7.1(1)}{\nano\farad}$.

\section{Calibration at PTB}
\label{Sec::PTB}

Two calibrations of both KATRIN dividers were carried out at the laboratory for high-voltage metrology (working group 2.32) of the PTB Braunschweig. All measurements were done in comparison to the MT100 \cite{Marx}, the standard divider of the PTB which is one of the most precise and most stable high voltage dividers in the world \cite{Divider_comparison}. Each measurement was done for both polarities and the uncertainties in this section are always given in units of standard deviations $\sigma$. The systematic uncertainties for the calibration of the 1818:1 and 3636:1 scale factors of the KATRIN dividers amount to \unit{2}{\ppm} ($k=2$) as stated in \cite{K35}. The calibration of the 100:1 scale factor was carried out at \unit{1}{\kV} directly with the Fluke 752A divider. Therefore its systematic uncertainty is \unit{1.2}{\ppm} ($k=2$) because it only depends on the transfer uncertainties of the digital voltmeters and the uncertainty of the reference divider (see table \ref{tab:M100_uncertainty}).

\begin{table}[!!!ht]
\begin{center}
\caption{Summary of the uncertainties of the calibration of the 100:1 scale factor at 1kV. The uncertainties are taken from the corresponding specifications. For the Fluke 752A reference divider the uncertainty in normal distributed and the coverage factor is k = 3. For both digital voltmeter the uncertainty relative to calibration standards over \unit{24}{\hour} was chosen. This is state in the manual of the Agilent 3458A with a rectangular distribution and for the Fluke 8508A with a normal distribution and a coverage factor of k = 2.}
\begin{tabular}{lccc} 
\hline
Instrument  &probability distribution & divisor & relative uncertainty (ppm) \\ 
\hline
\hline
Fluke 752A  &Normal (k = 3) & 3 & $0.5/3 = 0.17$ \\ 
Fluke 8508A  &Normal (k = 2)& 2 & $0.9/2 = 0.45$ \\ 
Agilent 3458A  &Rectangular & $\sqrt{3}$ &$0.56/\sqrt{3} = 0.32$\\ 
\hline
\multicolumn{3}{l}{combined standard uncertainty}& 0.58 \\ 
\hline 
\end{tabular} 
\label{tab:M100_uncertainty}
\end{center}
\end{table}
\medskip

\subsection*{Calibration phase 2009}

The first measurements of the K65 divider at the PTB were carried out from October to December 2009. During this phase the scale factors, the voltage dependence  and the temperature dependence was measured. Although the voltage divider provides an additional 566:1 scale factor, this one has not been investigated. The determined parameters are shown in table \ref{tab:PTB2009}.

\begin{table}[!!!htbp]
\caption{Summary of the calibration phase in 2009 at PTB. Listed are the scale factors determined at the given calibration voltage, their relative standard deviations over all measurements and their calibration uncertainties. The measurements were always done with voltages of both polarities. The temperature dependence was measured for the 1818:1 scale factor and temperature stability was found to be the same for all scale factors. The linear voltage dependence is given as mean value for the scale factors 1818:1 and 3636:1.}
\begin{center}
\begin{tabular}{cccc}
&&&\\
\hline 
Parameter & 100:1  & 1818:1  & 3636:1  \\ 
\hline 
scale factor  & \multirow{2}{*}{100.51484} & \multirow{2}{*}{1818.1096} & \multirow{2}{*}{3636.2743} \\ 
(at \unit{1}{\kV} for 100:1 else \unit{35}{\kV}) & & & \\
relative standard deviation & 1.2$\cdot10^{-7}$ & 6.8$\cdot10^{-7}$ & 7.0$\cdot10^{-7}$ \\ 
calibration uncertainty (k=2) & 1.2$\cdot10^{-6}$ & 2$\cdot10^{-6}$ & 2$\cdot10^{-6}$ \\
 &   &  &  \\
temperature dependence (18 to \unit{34}{\celsius}) & & $<1\cdot10^{-7}/\mathrm{K}$ & \\ 
temperature stability  & \multicolumn{3}{c}{$\pm$0.1K}\\ 
 &   &  &  \\
linear voltage  & & \multicolumn{2}{c}{\multirow{2}{*}{-1.9(2)$\cdot10^{-8}/$kV}}\\ 
dependence $\alpha'$  (8-32kV) & & \multicolumn{2}{c}{}\\
\hline 
\end{tabular}  
\label{tab:PTB2009}
\end{center}
\end{table}

\subsection*{Calibration phase 2011}

A second calibration phase was carried out from September to December 2011. The scale factors, the voltage dependence  and the warm-up behaviour was investigated during this phase.

\begin{figure}[!!!!!htbp]
\begin{center}
\includegraphics[width=0.95\linewidth]{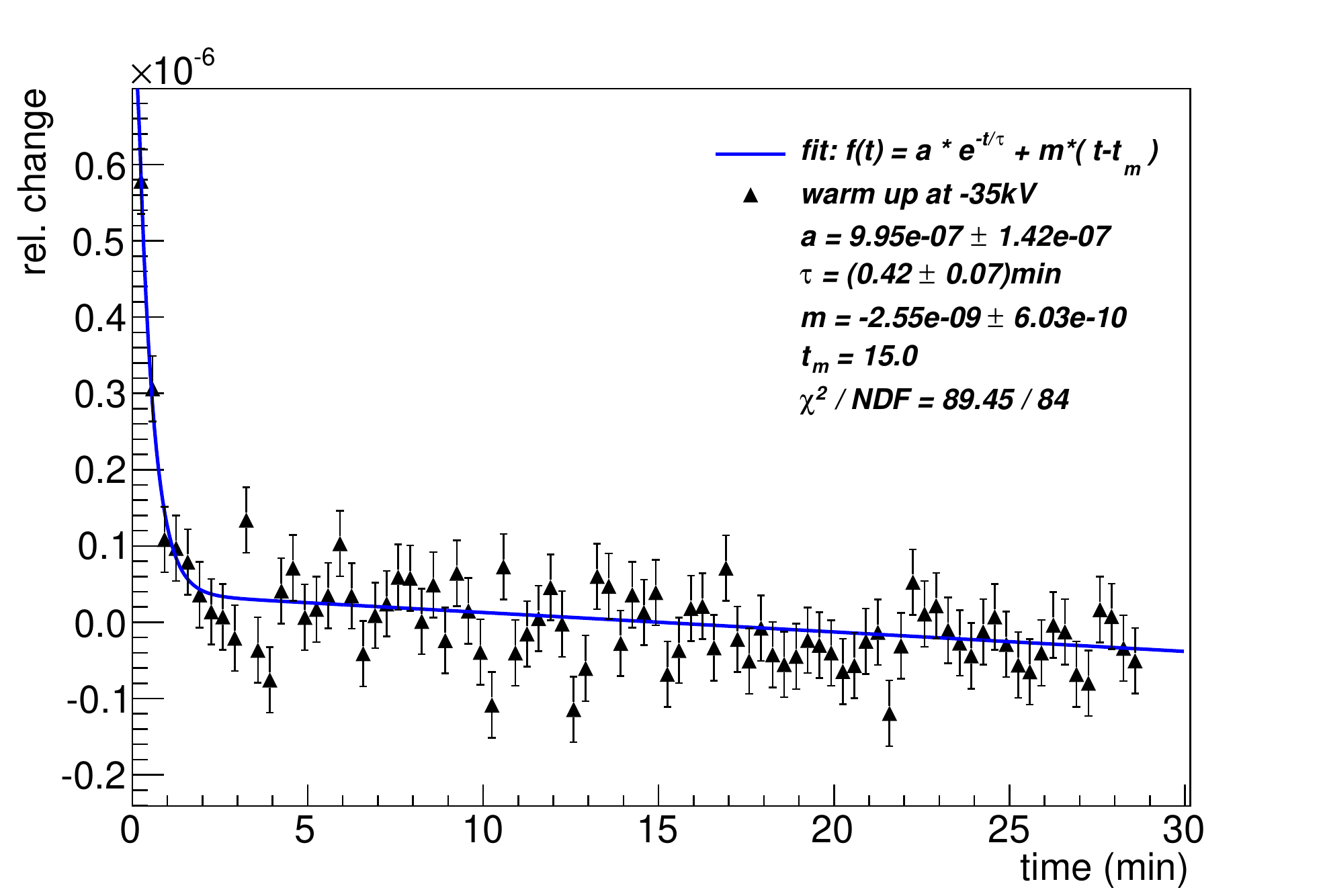} 
\caption{Relative change of the 1818:1 scale factor after applying \unit{-35}{\kilo\volt}. The scale factor measurement  in comparison to the scale factor of the MT100 
started with the application of \unit{-35}{\kilo\volt}. Before that the K65 divider was idle for at least \unit{30}{\min}. The plotted values are corrected for the measured warm-up behaviour of the MT100. The data set consists of the mean values of five independent measurements. The error bars represent the standard deviations between these five measurement series. The fit function was the sum of an exponential function and a straight line because a small linear drift was observed. This Fit yields a time constant of 0.42(7)\,min. The first measurement point was discarded and the fit was done for the whole measurement period of \unit{30}{\minute}. The small linear drift of $-2.6(6)\cdot10^{-9}/\mathrm{min}$ cannot be used to estimate a long-term drift. It can be caused from a slow process during the thermalization of the divider and is not observed over longer periods of time.}
\label{fig:Warm_up}
\end{center}
\end{figure}

For the warm-up behaviour the change in scale factor was measured over \unit{30}{\minute} after the high voltage was applied. To obtain a possible warm-up behaviour of the MT100 a dedicated measurement was made using the K65 as reference divider. This was done by connecting the MT100 via a vacuum circuit-breaker to the system. After the K65 was stabilized the voltage was ramped down fast, the MT100 was switched by the vacuum circuit-breaker to the system and the voltage was ramped up again. In all four measurements a slight deviation in the sub-ppm region was observed and all measurements were corrected for this warm-up behaviour. The warm-up behaviour of K65 was measured for the 1818:1 and 3636:1 scale factors at voltages of \unit{18.6}{\kilo\volt} and \unit{35}{\kilo\volt} with both polarities. As an example the warm-up behaviour at \unit{-35}{\kilo\volt} of the  1818:1 scale factor is shown in figure \ref{fig:Warm_up}. The measurements at \unit{18.6}{\kilo\volt} did not show any measurable warm-up deviation.

\begin{figure}[!!!!!htbp]
\begin{center}
\includegraphics[width=0.9\linewidth]{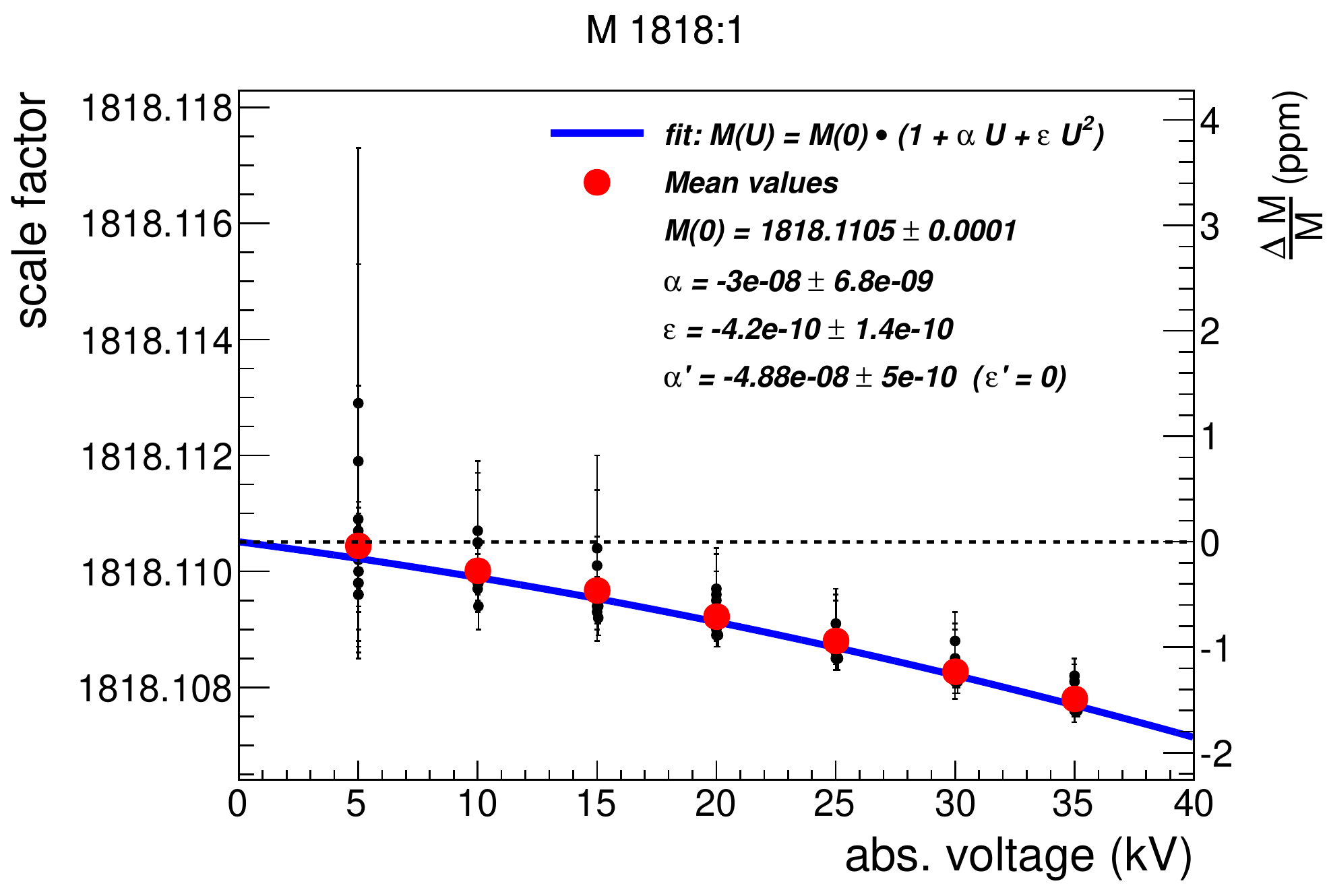} 
\includegraphics[width=0.9\linewidth]{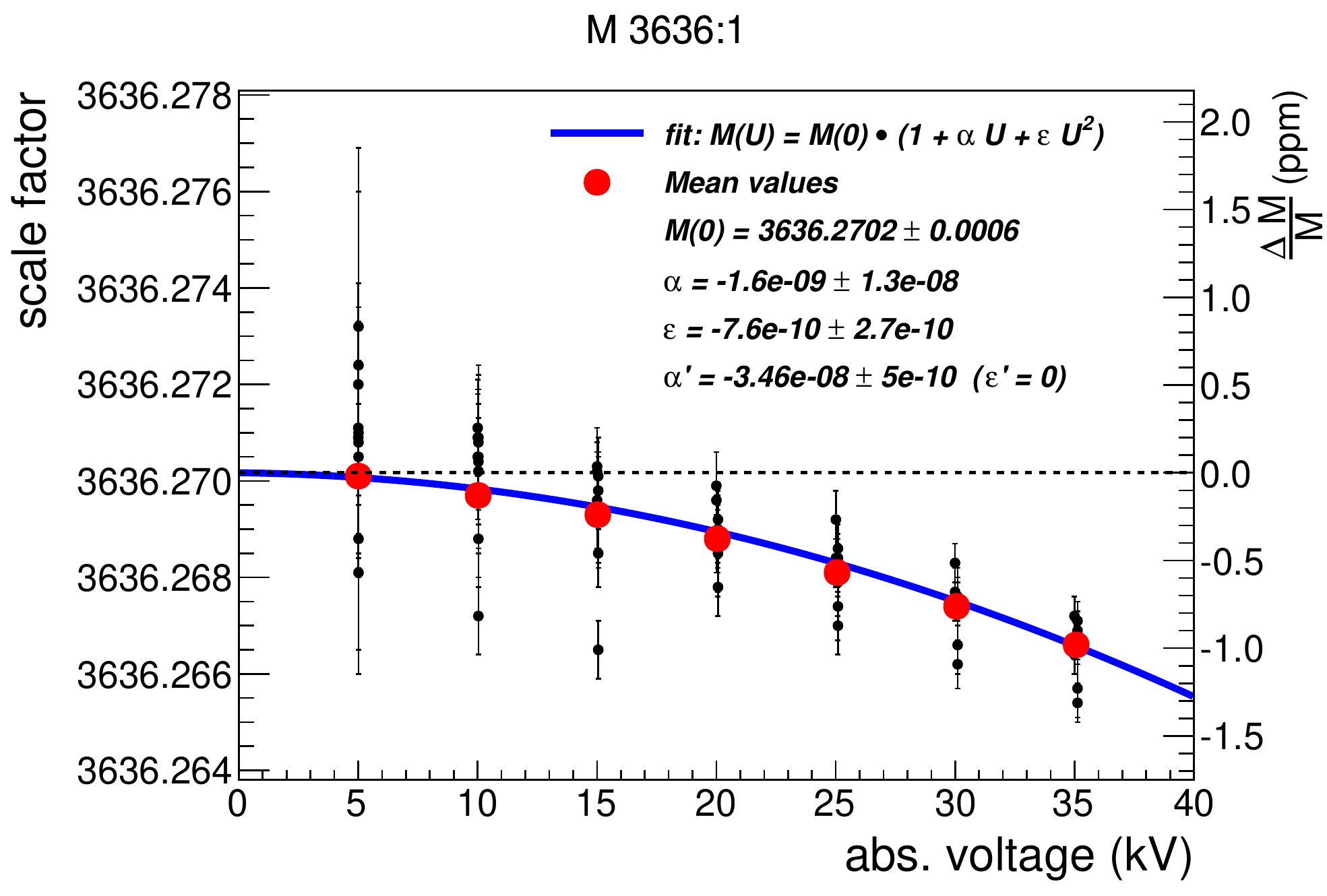}
\caption{Voltage dependence of the 1818:1 and 3636:1 scale factors.  The data sets were fitted by a 2$^{nd}$ order polynomial (blue line, coefficients $\alpha, \varepsilon$) and by a linear function (not plotted, coefficients $\alpha', \varepsilon'$). The relative change for both scale factors after a voltage step of \unit{35}{\kilo\volt} is below \unit{1.5}{\ppm}.}
\label{fig:Uabh}
\end{center}
\end{figure}

The voltage dependence of the divider was investigated by multiple measurements between \unit{5}{\kilo\volt} and \unit{35}{\kilo\volt} in steps of \unit{5}{\kilo\volt} for both polarities (see figure \ref{fig:Uabh}). These data sets were fitted by a 2$^{nd}$ order polynomial. The quadratic term is motivated by electric heating of the resistors and their temperature dependent resistance. The linear term is a combination of various effect. On the one hand leakage currents will go linearly with the applied voltage but also secondary effects of the heating of the resistors (like temperature dependent heat conduction) can cause a linear effect. Hence it is not possible to  quantify the leakage current effect without further investigations. In addition the data sets were fitted by a linear function in the range between \unit{10}{\kilo\volt} and \unit{35}{\kilo\volt} to compare these results to older ones. For a voltage step of \unit{35}{\kilo\volt} the relative change in scale factor is approximately \unit{1.5}{\ppm}.

To determine the long-term stability of the divider, a linear time dependence between the two calibrations in 2009 and 2011 was assumed. For all scale factors of the K65 divider the obtained long-term stability was measured to be less than \unit{0.1}{\ppm\per\Month}. 

\begin{table}[!!!htbp]
\caption{Summary of the calibration phase in 2011 at PTB. Listed are the scale factors for the applied calibration voltage, their relative standard deviations over all measurements and their calibration uncertainties. The temperature stability was found to be the same for all scale factors. The linear voltage dependence (measured for both polarities) is given for the 1818:1 and the 3636:1 scale factor. The warm-up deviation and time constant was measured for the 1818:1 and the 3636:1 scale factor and in both cases the deviation is less than \unit{1}{\ppm} and less than \unit{1}{\minute}.}
\begin{center}
\begin{tabular}{cccc}
&&&\\
\hline 
Parameter & 100:1  & 1818:1  & 3636:1  \\ 
\hline 
scale factor & \multirow{2}{*}{100.51479} & \multirow{2}{*}{1818.1078} & \multirow{2}{*}{3636.2668} \\ 
(at \unit{1}{\kV} for 100:1 else \unit{35}{\kV}) & & & \\
rel. standard deviation & 4.5$\cdot10^{-7}$ & 1.1$\cdot10^{-7}$ & 1.5$\cdot10^{-7}$ \\ 
calibration uncertainty (k=2) & 1.2$\cdot10^{-6}$ & 2$\cdot10^{-6}$ & 2$\cdot10^{-6}$ \\
 &   &  &  \\
temperature stability  & \multicolumn{3}{c}{$\pm$0.1K}\\ 
 &   &  &  \\
linear voltage  & & -4.88(5)$\cdot10^{-8}/$kV & -3.46(5)$\cdot10^{-8}/$kV\\ 
dependence $\alpha'$ (10-35kV) & \multicolumn{3}{c}{}\\
 &   &  &  \\
warm-up deviation & & \multicolumn{2}{c}{<1$\cdot10^{-6}$ (see figure \ref{fig:Warm_up})}\\
warm-up time constant & & \multicolumn{2}{c}{<1min (see figure \ref{fig:Warm_up})}\\
&   &  &  \\
stability (per month) & 2$\cdot10^{-8}$ & 4(3)$\cdot10^{-8}$ & 9(3)$\cdot10^{-8}$\\
\hline 
\end{tabular}  
\end{center}
\label{tab:PTB2011}
\end{table}

\section{Conclusion and outlook}

The K65 divider is a consequent further development of the K35 divider and has improved thermal characteristics and better long-term stability.
The concept of the TCR compensation by paired, pre-aged resistors in a temperature-stabilized vessel 
with guard-electrode system provides a high voltage divider in the ppm-class. It (over-)fulfils 
the requirements of the KATRIN experiment in all respects and hence enables a reduction of systematic uncertainties. As a transportable device the K65 could be used for other applications too.  

The usage of pre-aged resistors leads to a stability over time of the order of \unit{0.1}{\ppm\per\Month}. The linear voltage dependence of the various scale factors is in the sub-ppm region per kV and the influence by the warm-up drift of the resistors is very small as expected. 
After a warm-up phase of about one minute the sub-ppm accuracy is reached after a voltage step of \unit{35}{\kilo\volt}.
The voltage dependence can be described by a second order polynomial to describe different effects like leakage currents and the temperature coefficient of resistance of the resistors. The influence of leakage currents was already mentioned in \cite{YiLi_Divider} but the exact influence at the K65 can not be stated without further investigations. 

We see only little room to improve this type of high voltage divider with these kind of resistors even further. Probably one could  reduce the leakage currents even further by increasing the surface resistance of all insulators, e.g. by grooving the surfaces of the supports or by using materials with higher surface resistance. To improve the thermal dependence one could increase the resistance of the control divider and/or split it up, e.g. into four divider chains instead of one  to distribute the thermal load within the vessel more equally. 

The scale factors of a high voltage divider with such low thermal dependencies can be even calibrated with lower voltages still yielding a precision of a few ppm. This can be done by using the Fluke 752A reference divider and input voltages of \unit{1}{\kV} and less. Firstly the 100:1 scale factor is determined by applying \unit{1}{\kV} to the high voltage input. Secondly a voltage of  \unit{350}{\volt} is applied to the connector of the 100:1 scale factor. Using the 100:1 tap as input yields the same output voltage at the scale factors 1818:1 and 3636:1 as if \unit{35}{\kV} is applied to the high voltage input. 
Combining the scale factors measured in the two steps this  "low voltage calibration  technique" allows to determine the scale factors 1818:1 and 3636:1 with an uncertainty of less than \unit{5}{\ppm} using commercial available devices only. Details and results of this low voltage calibration will be presented in \cite{HV_Cal}.

\section*{Acknowledgment}
We would like to thank Mr. H. Faierstein, Mr. D. Sachau and Mr. F. Weise from the company VISHAY for the very constructive, successful and nice cooperation to make this ppm-class high voltage divider possible. 
This work was supported by the German Ministry for Education and Research BMBF under references 05A08PM1 and 05A11PM2.

\end{document}